\documentclass[a4paper,11pt]{article}
\usepackage[english]{babel}
\usepackage[utf8]{inputenc}
\usepackage[T1]{fontenc}
\usepackage{lmodern}
\usepackage{amsmath,amsfonts,amsthm,amssymb}
\usepackage{mathptmx}%
\usepackage{newtxtext,newtxmath} %
\usepackage[scaled=.95]{helvet} %
\usepackage[left=30mm,right=30mm,top=30mm,bottom=30mm,marginparwidth=17.5mm]{geometry}
\usepackage[hyphens]{url}
\usepackage{hyperref}
\usepackage{doi}
\usepackage{graphics}
\usepackage{graphicx}
\usepackage{color}
\usepackage{microtype}
\usepackage[table,xcdraw]{xcolor}
\usepackage[round]{natbib}
\usepackage{tikz}
\usepackage{amsbsy}
\usepackage{dsfont}
\usepackage{mathtools}
\usepackage{bm}
\usepackage{bbm}
\usepackage[font=small, skip=7pt]{caption}
\usepackage{subcaption}
\usepackage{float}
\usepackage{rotating}
\usepackage{flafter} %
\usepackage{titling}
\usepackage[auth-lg]{authblk} %
\usepackage{fancyhdr} %
\usepackage{appendix}
\usepackage{booktabs}
\usepackage{enumitem}
\usepackage{nicefrac}
\usepackage[capitalise,nameinlink,sort&compress]{cleveref} %

\usepackage{numprint}%
\npdecimalsign{.}
\npthousandsep{,}

\usetikzlibrary{arrows,shapes,trees,shapes.geometric}

\hypersetup{
    colorlinks,
    breaklinks,
    linkcolor={blue!60!black},
    citecolor={blue!60!black},
    urlcolor={blue!60!black},
    pdftitle={Extreme Conformal Prediction: Reliable Intervals for High-Impact Events},
    pdfauthor={Olivier C. Pasche},
    pdfkeywords={}
}

\setlist{topsep=0.75\topsep, partopsep=0.75\partopsep, itemsep=0.75\itemsep, parsep=0pt}

\theoremstyle{plain}

\newtheorem{theorem}{Theorem}[section]

\newtheorem{proposition}[theorem]{Proposition}

\theoremstyle{remark}

\newcommand{\R}{\mathbb{R}}

\newcommand{\Prob}{\mathbb{P}}

\newcommand{\abs}[1]{\left\lvert#1\right\rvert}
\DeclarePairedDelimiter{\ceil}{\lceil}{\rceil} %
\DeclarePairedDelimiter{\floor}{\lfloor}{\rfloor} 
\newcommand{\Ytest}{Y_{\text{test}}}
\newcommand{\Xtest}{\bm{X}_{\text{test}}}

\newcommand{\Stest}{S_{\text{test}}}
\newcommand{\ymin}{y_{\text{min}}}
\newcommand{\qGPD}{\hat{Q}^{\text{GPD}}}

\newcommand{\indep}{\perp\kern-5.5pt\perp}

\graphicspath{{fig/}}	

\let\LTXmaketitle\maketitle
\renewcommand{\maketitle}{{\bf\LTXmaketitle}} %

\title{Extreme Conformal Prediction: Reliable Intervals for High-Impact Events}
\newcommand{\shorttitle}{Extreme Conformal Prediction}

\author[1,2,*]{Olivier C. Pasche}
\author[2]{Henry Lam}
\author[1]{Sebastian Engelke}

\affil[1]{Research Institute for Statistics and Information Science, University of Geneva, Switzerland}
\affil[2]{Department of Industrial Engineering and Operations Research, Columbia University, New York, USA}
\affil[*]{Corresponding author: \normalfont{\texttt{olivier.pasche@unige.ch}}}

\date{}

\begin{document}

\pagestyle{fancy}
\fancyhead{} %
\fancyhead[C]{\small\sc\shorttitle}
\renewcommand{\headrulewidth}{0pt} %

\maketitle

\begin{abstract}
Conformal prediction is a popular method to construct prediction intervals with marginal coverage guarantees from black-box machine learning models. In applications with potentially high-impact events, such as flooding or financial crises, regulators often require very high confidence for such intervals. 
However, if the desired level of confidence is too large relative to the amount of data used for calibration, then classical conformal methods provide infinitely wide, thus, uninformative prediction intervals. 
In this paper, we propose a new method to overcome this limitation. We bridge extreme value statistics and conformal prediction to provide reliable and informative prediction intervals with high-confidence coverage, which can be constructed using any black-box extreme quantile regression method. 
A weighted version of our approach can account for nonstationary data.
The advantages of our extreme conformal prediction method are illustrated in a simulation study and in an application to flood risk forecasting.
\end{abstract}

{%
}

\section{Introduction}\label{s:intro}

Conformal prediction is a simple approach to producing prediction sets from any regression or classification model.
For a covariate vector $\bm{X}$ with values in $\mathcal{X}\subseteq\R^p$ and corresponding response variable $Y$, the goal of classical conformal prediction is to build a prediction set $\hat{C}(\bm{x})$ satisfying marginal coverage
\begin{equation}\label{e:mcov}
    \Prob\{\Ytest \in \hat{C}(\Xtest)\} \geq 1-\alpha, 
\end{equation}
for a desired confidence level $1-\alpha\in(0,1)$, for any new observations $(\Xtest, \Ytest)$.
To obtain such a prediction set, we assume that a prediction model $\hat f$ was fitted on a training data set from the distribution $\mathcal P$ of $(\bm{X}, Y)$, and that a new calibration data set $(\bm{X}_1, Y_1), \dots, (\bm{X}_{n_c}, Y_{n_c})$ of $n_c$ observations from the same distribution is available.
For a specific nonconformity score function $s:\mathcal X \times \mathbb R \to \mathbb R$ that may depend on $\hat{f}$ and acts on the predictors and responses, consider the calibration scores $S_i = s(\bm{X}_i, Y_i)$ for $i=1,\dots, n_c$. 
For some level $\alpha\in(0,1)$, denoting by $\hat{q}_\alpha$ the $\{\ceil*{(n_c+1)(1-\alpha)}/n_c\}$-quantile of the scores $S_1,\dots, S_{n_c}$, the prediction set 
\begin{equation}\label{e:conf_int}
   \hat{C}(\bm{x}) = \{y : s(\bm{x}, y) \leq \hat{q}_\alpha\}
\end{equation}
has the desired $1-\alpha$ coverage. 
That is, it satisfies~\cref{e:mcov} for any new test observation $(\Xtest,\Ytest)$ from the same distribution $\mathcal P$, under a remarkably weak exchangeability assumption. 
This well-established framework is the so-called split conformal approach \citep{SplitConformalMargProof,Lei18ConformalPred}. %
More generally, originally started in \citet{vovk1999ml,vovk2005algorithmic}, the conformalization idea that leverages quantile-based construction of prediction sets elicits a range of variants, with focus on optimal data usage and applying to different problems, including jackknife+ \citep{alaa2020discriminative,barber2021predictive}, cross-conformal prediction \citep{vovk2015cross} and ensemble-based approaches \citep{kim2020predictive,gupta2022nested}. 
In combination with machine learning methods, it can efficiently capture predictive uncertainty~\citep{shafervovk2008tutorial,Zhou25ConformSurveyML}, and provide intervals with highly adaptive lengths~\citep{ConformalizedQR}.
Extensions to nonexchangeable data are also well-studied~\citep{OliveiraConformNonexchError}.
In particular, weighted conformal methods can account for distribution shifts and drifts~\citep{TibshiraniCoformShift,Barber2023nonstatConform}.

Conformal prediction intervals are widely used for confidence levels $1-\alpha$ of moderate value relative to the sample size $n_c$ (e.g., $90\%$ and $n_c = 1000$), where enough of the calibration scores are above this quantile.
In many applications, however, test points $\Ytest \notin \hat{C}(\Xtest)$ that fall outside of the prediction set correspond to a high-impact event with serious consequences for the environment, human lives or the economy. Examples for such risk-sensitive applications are the protection of cities and energy infrastructure from flooding \citep{kee2009,AsadiDanube} or the financial reserves of banks and insurance companies \citep{zhou2019, dup2022}.
In these cases, much larger values of confidence $1-\alpha$, close to one, will be required, sometimes even by law.
Classical methods from conformal prediction fail for those requirements, since the quantile $\hat{q}_\alpha$ as defined above is not a useful estimator when the level $\alpha < 1/(n_c+1)$. 
Indeed, in this case, less than one observation then exceeds the $(1-\alpha)$-quantile on average, and $\hat{q}_\alpha$ is infinite (or ill-defined). 
Even for slightly larger values $\alpha$ close to that limit, the variance of $\hat{q}_\alpha$ can be huge.

Extreme value theory provides statistical tools for accurate estimation beyond the data range \citep{DF2006}. 
The tools have proven successful for improving extrapolation properties of machine learning methods in regression \citep{de2022regression, huet2024regressionextremeregions, bur2024}, classification \citep{jalalzai:clemencon:sabourin:2018}, and generative methods \citep{bou2022}.

In this paper, we propose a new methodology that bridges the wide applicability of conformal prediction with extrapolation tools from extreme value statistics to construct reliable prediction sets for high-impact events. 
In a first step, in order to obtain a good pretrained model $\hat f$ beyond the data range, we rely on flexible machine learning methods from extreme quantile regression \citep{Velthoenetal2019, erf, Pasche2024,RichardsNNfire}.
Second, we rely on the classical and theoretically justified peaks-over-threshold approach, which consists of using the generalised Pareto distribution (GPD) to extrapolate, for example, quantile estimates beyond the range of empirical observations~\citep{GPDBalkema,GPDPickands}.
For a confidence level $1-\alpha$ close to one, we leverage the GPD fitted to the calibration scores $S_1,\dots, S_{n_c}$ to obtain a reliable estimate of ${q}_\alpha$ beyond the calibration data. 
The resulting extreme conformal prediction intervals have better properties compared to those from the classical empirical approach for large confidence requirements. 
In a simulation study, we show that our method improves existing approaches in terms of better coverage, in the sense of \cref{e:mcov}, and of informativeness of the prediction interval.
We also consider a weighted version of our extreme conformal approach to deal with nonexchangeable data, and discuss its usage with several classical types of conformal procedures.

The advantages of our approach are illustrated in an application to flood risk prediction. 
Using several of the flexible machine learning methods as base predictions, it provides high-confidence one-day-ahead interval forecasts of the conditional range for water flow. 
We show that using conformal prediction intervals based on extreme value theory improves the coverage of the classical method, which either yields uninformative intervals or exhibits undercoverage and, therefore, seriously underestimates the risk of high-impact events.

\section{Background on conformal prediction}\label{s:background}

\subsection{Split conformal prediction}\label{ss:classicalsplitconform}
As in the introduction, let $\bm{X}$ be a covariate vector taking values in $\mathcal{X}\subseteq\R^p$, and $Y$ the response random variable of interest. 
We will consider, here and in the sequel, the regression case where the response $Y$ is real-valued in $\R$.
We also suppose that we have access to a calibration set of $n_c$ observations $(\bm{X}_1, Y_1), \dots, (\bm{X}_{n_c}, Y_{n_c})$.
Classical split conformal prediction builds prediction intervals (PIs) as in \cref{e:conf_int} that have desired coverage under very weak assumptions. 
Specifically, if the joint distribution of the calibration set and the test point $(\Xtest, \Ytest)$ is exchangeable, then the unconditional coverage guarantee~\cref{e:mcov} holds \citep{SplitConformalMargProof}.

Importantly, the probability measure in \cref{e:mcov} is with respect to the randomness in the calibration set jointly with the test point, that is, $\{(\bm{X}_1,Y_1),\ldots,(\bm{X}_{n_c},Y_{n_c}), (\Xtest, \Ytest)\}$. In fact, when conditioning on the calibration set, the distribution of the coverage is a beta distribution, that is,
\begin{equation}\label{e:mcovcalbeta}
    \Prob\left[\Ytest \in \hat{C}(\Xtest)\mid \{(\bm{X}_i,Y_i)\}_{i=1}^{n_c}\right] \sim {\rm Beta}(n_c+1-l,\, l), \quad l := \floor*{(n_c + 1)\alpha}.
\end{equation}
Running the conformal prediction twice on different calibration sets, therefore yields PIs with different coverage probabilities. The guarantee in \cref{e:mcov} says that when averaging out the calibration set, the coverage is at least $1-\alpha$.

Furthermore, the marginal coverage property in \cref{e:mcov} only guarantees ``overall'' marginal coverage of the prediction set $\hat{C}(\bm{x})$, but does not imply the conditional coverage property
\begin{equation}\label{e:ccov}
    \Prob\{\Ytest \in \hat{C}(\bm{x})\mid \Xtest=\bm{x}\} \geq 1-\alpha, \quad \forall \bm{x}\in\mathcal{X}.
\end{equation}
The latter is generally impossible to guarantee in such a general setting. How close $\hat{C}(\bm{x})$ is to satisfying the conditional coverage property depends on the quality of the given pretrained model. For example, for the conformalized quantile regression approach described in \cref{ss:CQR}, it depends on the accuracy of the initial quantile regression model $\hat{Q}_{\alpha/2}(\bm{x}), \hat{Q}_{1-\alpha/2}(\bm{x})$.

\subsection{Conformalized quantile regression}\label{ss:CQR}

Conformalized quantile regression, first proposed by~\citet{ConformalizedQR} and also described in~\citet{ConformalIntrRev}, is one of the most popular conformal methods, in particular thanks to its ability to provide varying-length prediction intervals with competitive adaptivity. 
Suppose that we have access to a black-box quantile regression model trained to estimate the conditional quantiles $\hat{Q}_{\alpha/2}(\bm{x})$ and $\hat{Q}_{1-\alpha/2}(\bm{x})$ of $Y$, given $\bm{X}=\bm{x}$, at probability levels $\alpha/2$ and $1-\alpha/2$, respectively. Then, conformalized quantile regression uses the score function 
\begin{equation}\label{e:scorebilat}
    s(\bm{x},y):=\max\{\hat{Q}_{\alpha/2}(\bm{x}) - y,\; y - \hat{Q}_{1-\alpha/2}(\bm{x})\}.
\end{equation}
Following the general procedure described in the introduction, this leads to the final prediction set in \cref{e:conf_int} being the interval
\begin{equation}\label{e:PI_twosided}
    \hat{C}(\bm{x}) = \left[\hat{Q}_{\alpha/2}(\bm{x}) - \hat{q}_\alpha,\; \hat{Q}_{1-\alpha/2}(\bm{x}) + \hat{q}_\alpha\right],
\end{equation}
where $\hat{q}_\alpha$ is the empirical $\{\ceil*{(n_c+1)(1-\alpha)}/n_c\}$-quantile of the calibration scores $S_1, \ldots, S_{n_c}$, i.e.\ the order statistic $S_{(\ceil*{(n_c+1)(1-\alpha)})}$.
As an equivalent definition, $\hat{q}_\alpha$ equals the empirical $(1-\alpha)$-quantile of $\{S_1, \ldots, S_{n_c}\}\cup\{+\infty\}$, the calibration score sample augmented with a point at infinity.

Intuitively, the procedure either widens (with a positive $\hat{q}_\alpha$) or narrows (with a negative $\hat{q}_\alpha$) the initial interval $\left[\hat{Q}_{\alpha/2}(\bm{x}), \hat{Q}_{1-\alpha/2}(\bm{x})\right]$ so that it covers $\ceil*{(n_c+1)(1-\alpha)}$ of the $n_c$ calibration observations. Note that the resulting prediction intervals satisfy the marginal coverage~\cref{e:mcov}, but there is no guarantee that the conditional coverage in~\cref{e:ccov} is satisfied. In fact, the more accurate the initial quantile regression models $\hat{Q}_{\alpha/2}(\bm{x})$ and $\hat{Q}_{1-\alpha/2}(\bm{x})$ are, the better the conditional coverage will be.

\subsection{Limitation for extreme confidence levels}\label{ss:limitations}

For high-impact events, regulators often require predictions with very high coverage probabilities to ensure that protective infrastructures or measures are sufficient. In particular, in such risk-sensitive applications, the level $\alpha$ in \cref{e:mcov} is typically close to $0$ and may satisfy $\alpha < 1/(n_c+1)$. This is generally referred to as an extreme confidence or probability level since, on average, there is less than one observation above the $(1-\alpha)$-quantile in a sample of size $n_c$. Note that the size $n_c$ of the calibration set is typically fairly small, since these data cannot be used for model fitting, often resulting in extreme scenarios even for relatively moderate levels of $\alpha$.

The classical construction of the conformal prediction intervals described in the introduction requires the computation of 
$\hat{q}_\alpha$, the empirical $\{\ceil*{(n_c+1)(1-\alpha)}/n_c\}$-quantile of the calibration set scores. 
For extreme confidence levels $1-\alpha$, this quantile level 
\begin{equation*}
    \ceil*{(n_c+1)(1-\alpha)}/n_c > \ceil*{n_c}/n_c = 1,
\end{equation*}
in which case $\hat{q}_\alpha$ is ill-defined and, by convention, set to infinity \citep{ConformalizedQR,ConformalIntrRev}. 
This results in degenerate trivial prediction intervals $\hat{C}(\bm{x}) = \left(-\infty, \infty\right)$, for all $\bm{x}\in\mathcal{X}$. 
Although this interval satisfies the coverage~\cref{e:mcov}, it is of no practical utility.

\section{Extreme conformal prediction}\label{s:method}

We propose an approach based on extreme value statistics to construct nondegenerate conformal prediction intervals at extreme confidence levels $1-\alpha>n_c/(n_c+1)$. %
Similarly to classical conformalized quantile regression~\citep{ConformalizedQR}, our method requires two steps:
\begin{enumerate}
    \item fitting a quantile regression model at level $1-\alpha$ on a training data set of size $n$;
    \item calibrating based on the scores $S_1,\dots, S_{n_c}$ on an independent data set. 
\end{enumerate}
For extreme confidence levels, both steps typically require extrapolation beyond the data range. Indeed, if $\alpha$ is close to 0, and in particular if $\alpha < 1/(n+1)$ is also extreme in the training data, then usual quantile regression will not be accurate. Instead, extreme quantile regression methods should be used. There is large literature on such methods based on linear models \citep{chernozhukov2005}, additive models \citep{ExGAM,ExGAM2,de2022bayesianLRtails}, or more flexible machine learning models such as gradient boosting \citep{gbex,Kohfireboosting}, random forest~\citep{erf} or neural networks \citep{Pasche2024,RichardsNNfire,AlloucheNN}. Importantly, the model in step~1 can be a black-box, in the sense that we do not require theoretical guarantees. 
We discuss the extrapolation for step~2 in \cref{ss:extrconform} and come back to examples of extreme quantile regression models in \cref{s:sim,s:appl}. 
Furthermore, we discuss extensions of our approach to nonexchangeable data and alternative conformal procedures in \cref{ss:extensions,apx:extensionsdetails}.

\subsection{Single-sided prediction intervals}\label{ss:singlesidedPI}

Extreme conformal prediction intervals are most relevant in cases where very large values of the response variable $Y$ lead to severe negative impacts. 
In such cases, reliable prediction intervals for $Y\mid\bm{X}=\bm{x}$, which contain the realisation of $Y$ with very high marginal probability, are a crucial forecasting tool. 
They can be used to determine whether a dangerous level of the response could potentially be reached. This also allows the reduction of false negatives (e.g., in the form of missing warnings) that can be critical to the system.
In those risk assessment scenarios, it is often a single of the two tail directions which is of risk. 
This is the case for the risk of flooding discussed in \cref{s:appl} but also, for instance, for high temperatures in dry areas at risk of wildfires, and financial asset returns at risk of large losses.
Without loss of generality, we, thus, suppose that one is interested in single-sided prediction intervals; two-sided intervals can be constructed analogously. 

The classical procedure, yielding two-sided intervals, can be adapted to obtain single-sided prediction intervals by using the score function
\begin{equation}\label{e:scoreunilat}
    s(\bm{x},y):= y - \hat{Q}_{1-\alpha}(\bm{x}),
\end{equation}
instead of \cref{e:scorebilat}, where $\hat{Q}_{1-\alpha}(\bm{x})$ is the pretrained quantile regression model at level $1-\alpha$.
Let $\ymin\in\R\cup\{-\infty\}$ be the lower endpoint of the distribution of $Y$ (or of the conditional distribution $Y\mid\bm{X}=\bm{x}$, if known). Then, following the usual procedure, the resulting interval is 
\begin{equation}\label{e:PI_singlesided}
    \hat{C}(\bm{x}) = \left(\ymin,\; \hat{Q}_{1-\alpha}(\bm{x}) + \hat{q}_\alpha\right],
\end{equation}
where $\hat{q}_\alpha$ is, in classical conformal prediction, the $\{\ceil*{(n_c+1)(1-\alpha)}/n_c\}$-quantile of the calibration scores $S_1, \ldots, S_{n_c}$, i.e., the order statistic $S_{(\ceil*{(n_c+1)(1-\alpha)})}$.

\subsection{Calibrative extrapolation}\label{ss:extrconform}

As discussed in Section~\ref{ss:limitations}, when an extreme confidence level $1-\alpha > n_c/(n_c + 1)$ is required, using order statistics to estimate $\hat{q}_\alpha$ would lead to degenerate intervals.
Therefore, an alternative approach is needed to estimate a finite value $\hat{q}^e_\alpha$ from the calibration set such that 
\begin{equation}\label{e:goalExConf}
    \Prob\left(\Stest \leq \hat{q}^e_\alpha\right)\geq 1-\alpha.
\end{equation}
Substituting $\hat{q}^e_\alpha$ for $\hat{q}_\alpha$ in \cref{e:conf_int} (or in \cref{e:PI_twosided,e:PI_singlesided}) would, then, yield nondegenerate prediction intervals that satisfy the marginal coverage guarantee in~\cref{e:mcov}. 
We propose to rely on the classical peaks-over-threshold methodology from extreme value theory to find such a quantile estimate. 
The tail of the distribution of the calibration score $S:=s(\bm{X},Y)$ can be approximated by the generalized Pareto distribution (GPD) above a high threshold $u$ by
\begin{align}\label{e:GPD}
\Prob(S > y) = \Prob(S > u) \Prob(S>y \mid S>u) \approx \Prob(S > u) \left\{1 + \xi \frac{y-u}{\sigma(u)}\right\}_+^{-1/\xi},\quad y \geq u, %
\end{align}
where $\xi \in \mathbb R$ and $\sigma(u)>0$ are the shape and scale parameters
and $u$ is an intermediate threshold. 
Under very mild assumptions on the distribution $F_S$ of $S$, this approximation is theoretically justified as $u$ tends to the upper endpoint of $F_S$ ~\citep{GPDBalkema,GPDPickands}.
In practice, $u$ is typically chosen as the empirical $\tau_0$-quantile $\hat Q_{\tau_0}$ of $S$, for $\tau_0 = (1-k/n_c)$ and some $k<n_c$. The tuning parameter $k$ is the number of exceedances used for
estimation of the parameters $\sigma$ and $\xi$, for instance, by maximum likelihood.
Quantiles of $S$ can then be estimated at probability levels beyond the data range using the approximation 
\begin{equation}\label{e:gpdquant}
 \qGPD_{\tilde{\tau}}:= \hat Q_{\tau_0} + \frac{\hat{\sigma}}{\hat{\xi}} \left\{\left(\dfrac{1 - \tau_0}{1-\tilde{\tau}}\right)^{\hat{\xi}} -1 \right\},\quad \tilde{\tau} >\tau_0. %
\end{equation}
Theoretical guarantees of these estimators typically require that $k\to\infty$ and $k/n_c\to 0$ as $n_c\to\infty$ to ensure the correct tradeoff between bias and variance; see \cite{DF2006} for more details.

Asymptotically, as $n_c\to\infty$ and under additional second-order conditions, using $\hat{q}^e_\alpha:=\qGPD_{1-\alpha}$ would satisfy \cref{e:goalExConf} with equality (in a suitable limiting sense as $\tau_0 \to 1$). 
But, since the calibration sample is finite and the level $\tau_0$ fixed, $\qGPD_{1-\alpha}$ can underestimate the true quantile due to estimation and approximation biases, respectively \citep{Roodman2018,BMvsPOT,ZederPasche23}. 
We, therefore, follow here a more conservative approach. 
Alternatively to choosing $\hat{q}^e_\alpha$ as $\qGPD_{1-\alpha}$,
we may use the upper endpoint of a $(1-\alpha_2)$-confidence interval for ${F}^{-1}_{S}(1-\alpha_1)$, the $(1-\alpha_1)$-quantile of the calibration scores $S$, for two suitable levels $\alpha_1,\alpha_2\in(0,1)$. The following proposition shows that, if this confidence interval has correct coverage, then the resulting extreme conformal prediction interval satisfies~\cref{e:mcov}.  

\begin{proposition}\label{prop:CIcovr}
Let $\alpha_1,\alpha_2\in(0,1)$, and $[L_Q,U_Q]$ be a $(1-\alpha_2)$-confidence interval for ${F}^{-1}_{S}(1-\alpha_1)$, the $(1-\alpha_1)$-quantile of the calibration scores $S$. If the confidence interval has correct coverage, i.e.~$\Prob\{L_Q\leq {F}^{-1}_{S}(1-\alpha_1)\leq U_Q\}\geq 1-\alpha_2$, and if
\begin{align}\label{alpha_ineq}
    1-\alpha\leq (1-\alpha_1)(1-\alpha_2),
\end{align}%
then 
$\Prob\left(\Stest \leq U_Q\right)\geq 1-\alpha$. 
That is, 
\begin{equation}\label{extr_covr}
   \Prob\{\Ytest \in \hat{C}^e(\Xtest)\} \geq 1-\alpha, \quad \text{where} \quad \hat{C}^e(\bm{x}) := \{y : s(\bm{x}, y) \leq U_Q\}.
\end{equation}
\end{proposition}

The proposition applies more broadly to other types of base predictions and nonconformity score functions than to those of conformalized quantile regression. 
Its proof is presented in \cref{apx:proof_covr}.

A natural choice for $\alpha_1$ and $\alpha_2$ that satisfies~\cref{alpha_ineq} with equality is $\alpha_1=\alpha_2=1-(1-\alpha)^{1/2}$, which is analogous to the {Šidák} correction~\citep{Sidak1967}.
Another notable choice is $\alpha_1=\alpha_2=\alpha/2$, which is analogous to a Bonferroni correction~\citep{bonferroni1936}. Although the latter is, in this case, overconservative, the difference is negligible for small values of $\alpha$.

There are several well-studied approaches for obtaining extreme quantile confidence intervals~(CI) using the GPD approximation \citep{IExtr,DavisonBootstrap,DavisonBootstrap2,HaanZhouExtrBoot}, including the profile likelihood, the bootstrap and the normal-approximation delta method. 
The profile-likelihood CI typically represents the inherently asymmetrical uncertainty best and yields the most conservative CI upper endpoint. 
For very small $\alpha$ and small sample sizes, it sometimes suffers from numerical difficulties for estimating the CI's upper endpoint. 
They arise as the derivative of the profile log-likelihood can get close to zero on its right-hand side, which sometimes makes finding the crosspoint between the profile curve and the chi-square confidence line numerically difficult.
Slightly varying the value of the GPD threshold can sometimes solve the issue.
The bootstrap approach comes in different variations, both for the sampling step (nonparametric, parametric) and for the aggregation step (basic, percentile, normal, etc.). 
We here consider the nonparametric bootstrap with percentile aggregation, as it is the most commonly used. 
The bootstrap can give reliable confidence intervals, but, compared to the profile-likelihood approach, its upper endpoint estimates might not always be conservative enough. 
Finally, the delta-method CI is computationally less expensive than the other two alternatives, but it provides intervals that are symmetric around the quantile estimate, which is not realistic for large quantiles. Moreover, it can also suffer from numerical instability issues, due to its matrix inversion step, and fail to yield meaningful CIs. 
These alternative extreme CI methods are further discussed and compared in \cref{s:sim}.

To summarise, when given a high confidence level $1-\alpha$, above or close to $n_c/(n_c+1)$, our proposed extreme conformal prediction interval takes the form 
\begin{equation}\label{e:extremePI}
    \hat{C}^e(\bm{x}) = \left(\ymin,\; \hat{Q}^e_{1-\alpha}(\bm{x}) + \hat{q}^e_\alpha\right],
\end{equation}
where $\hat{Q}^e_{1-\alpha}(\cdot)$ is a prefitted extreme quantile regression model and $\hat{q}^e_\alpha$ is the upper endpoint of an appropriate GPD profile-likelihood confidence interval for a high quantile of the calibration nonconformity scores, as discussed above.
If a two-sided extreme PI is preferred, one can use the classical nonconformity scores from \cref{e:scorebilat}, instead of \cref{e:scoreunilat}, to obtain
\begin{equation}\label{e:extremePIbilat}
    \hat{C}^e(\bm{x}) = \left[\hat{Q}^e_{\alpha/2}(\bm{x}) - \hat{q}^e_\alpha,\; \hat{Q}^e_{1-\alpha/2}(\bm{x}) + \hat{q}^e_\alpha\right],
\end{equation}
where a second prefitted extreme quantile regression model $\hat{Q}^e_{\alpha/2}(\cdot)$ is required for extrapolating the lower-tail conditional quantiles.

\subsection{Extensions to other conformal approaches and nonexchangeable data}\label{ss:extensions}

We introduce our extreme conformal prediction approach as a variant of the well-established split conformalized quantile regression, as its use of base quantile predictions naturally results in varying-length intervals, which, in addition to valid marginal coverage, can also achieve good conditional coverage~\citep{ConformalizedQR,ConformalIntrRev}, and as the split procedure is significantly less computationally costly than full-conformal or $k$-fold alternatives.
Nevertheless, \cref{prop:CIcovr} applies more broadly, and our approach is, in principle, adaptable to other conformal approaches, including different base regression models and nonconformity scores, the alternative full-conformal and $k$-fold procedures, and weighted methods for nonexchangeable data.

The extension to different base predictive models and scores is the most straightforward, as the extreme conformalization described in \cref{ss:extrconform} is agnostic to their definitions.
This includes the classical split-conformal approach, using a conditional-mean base regression model $\hat{\mu}(\bm{x})$, instead of the quantile predictions $\hat{Q}_{\alpha/2}(\bm{x}), \hat{Q}_{1-\alpha/2}(\bm{x})$ \citep{SplitConformalMargProof,papadopoulos2008inductive,Lei18ConformalPred}, 
and its heteroscedastic variants \citep{papadopoulos2008normalizedCoform,papadopoulos2011regressionknn,Lei18ConformalPred}.
Although these alternatives would not require extreme quantile regression, they either yield fixed-length PIs or tend to underestimate conditional variability~\citep{ConformalizedQR}.
The extension to full-conformal procedures \citep{vovk1999ml,vovk2005algorithmic,shafervovk2008tutorial} is also straightforward, but requires repeating both the predictive model fit and our extreme conformalization from \cref{ss:extrconform} for a dense grid of potential response values.
Although it makes a more efficient use of the data than the split variant, it has an extremely high computational cost.
The same applies, but less extremely, to $k$-fold approaches, such as Jackknife+/CV+ and cross-conformal prediction \citep{barber2021predictive,vovk2015cross}.
The extensions of our method to these alternative conformalisation approaches are described in more detail in \cref{apx:extensionsdetails}.

Weighted conformal approaches allow for relaxing the exchangeability assumption to account, for example, for distribution shifts or drifts. %
In those approaches, each calibration observation is assigned a weight $w_i\in[0,1]$, $i=1,\ldots,n_c$, which reflects its similarity to the test observation, or more generally, the similarity between its score $S_i$ and the test score $\Stest$.
The classical empirical quantile of the calibration scores $\hat{q}_\alpha$ is then replaced by a weighted sample quantile. 
More precisely, it is redefined as the $(1-\alpha)$-quantile of the weighted empirical distribution $\bar{w}^n\sum_{i=1}^{n_c} w_i \delta_{s_i}+\bar{w}^n\delta_{\infty}$, where $\bar{w}^n=(\sum_{i=1}^{n_c} w_i + 1)^{-1}$ and $\delta_x$ is the Dirac measure at $x$ \citep{Barber2023nonstatConform,TibshiraniCoformShift}. 
As our proposed $\hat{q}^e_\alpha$ from \cref{ss:extrconform} is, for extreme confidence levels, based on likelihood inference, the natural analogous extension to nonexchangeable data would be to rely on weighted likelihood inference instead. 
Using the sample weights $w_i\in[0,1]$, $i=1,\ldots,n_c$, the procedure would remain the same as in \cref{ss:extrconform}, only using the weighted  
\begin{equation}\label{e:wgpdll}
    \ell^{\text{GPD}}_w(\sigma,\xi) = - \log(\sigma) \sum_{i=1}^{n_c} w_i - \left(\frac{1}{\xi}+1\right) \sum_{i=1}^{n_c} w_i \log\left(1 + \xi \frac{s_i - u}{\sigma}\right)_+, 
\end{equation}
instead of the classical GPD log-likelihood, to infer the $(1-\alpha_2)$-CI endpoint for ${F}^{-1}_{\Stest}(1-\alpha_1)$.
This weighted alternative should achieve a similar effect as classical weighted conformal prediction methods: proportionally use the calibration scores most similar to the test point during conformalization, to correct for nonexchangeable distribution drifts and shifts. 
This weighted extension is further discussed and applied in \cref{s:appl}.

\section{Simulation study}\label{s:sim}

\subsection{Experimental setup}\label{ss:simsetup}
To assess the different conformalization methods, we perform a simulation study with several scenarios.
The data is generated from
\begin{equation}\label{e:simsdata}
\begin{cases}
\bm{X}  \sim \text{Unif}\left([-1, 1]^{10}\right),\\
Y \mid \bm{X} = \bm{x} \sim \sigma(\bm{x})\cdot \varepsilon_Y, %
\end{cases}
\end{equation}
with $\sigma(\bm{x}): = 1 + 6 \phi(x_1, x_2)$, where $\phi$ is the bivariate Gaussian density with correlation $0.9$.
We consider two scenarios for the noise variable $\varepsilon_Y$: a heavy-tailed Student $t$ distribution $t_{\alpha(\bm{x})}$, with covariate-dependent tail index $\alpha(\bm{x}) = 1/\xi(\bm{x}): = 7\cdot\{1 + \exp(4x_1 + 1.2)\}^{-1} + 3$, and a light-tailed Gaussian $\text{N}(0,1)$ distribution. 
The former choice corresponds exactly to the generating process used in two extreme quantile regression benchmark studies \citep{gbex,Pasche2024}.

We consider several sizes for the calibration sets, with $n_c\in\{10^3, 10^{3.5}, 10^4\}$ observations.
For each calibration size, we repeat the experiments 100 times. 
We consider extreme PI confidence levels $1-\alpha$, with $\alpha\in\{10^{-3},10^{-3.5},10^{-4},10^{-4.5},10^{-5}\}$. 
We consider three choices for the base quantile predictions $\hat{Q}^e_{1-\alpha}(\cdot)$: the conditional-quantile ground truth and two different pretrained extreme quantile regression models.
The ground-truth choice aims at assessing the methods with ideal initial predictions. As all conformalization methods are translation invariant, adding first-order bias to the pretrained model would always lead to the same final PIs. 
For the first pretrained extreme quantile predictions, we use the extreme quantile regression neural networks (EQRN) model, as it performed best on this benchmark dataset~\citep{Pasche2024}, aiming at assessing our procedure with accurate but realistic initial predictions. The EQRN model is pretrained on \numprint{5000} observations generated from~\cref{e:simsdata}. Its hyperparameters and architecture were selected based on validation GPD deviance with a grid search.
Lastly, to investigate the performance of our extreme conformalization procedure for a poorly-performing model, which could happen in practice due to poor historical modelling choices, we also consider a linear GPD quantile model as base predictions. 
More precisely, we use linear quantile regression to obtain a conditional threshold $\hat{u}(\bm{x})$, and model the exceedances with a GPD having a linear parametrization $\sigma=\sigma_0+\sigma_1x_1+\sigma_2x_2$ for its scale parameter.
Although having the GPD extrapolation potential, its linear parametrization is a bad fit for the highly nonlinear dependence of $Y$ on $\bm{X}$.
Experiments with Gaussian-noise data were only performed with the ground-truth base predictions. 

For each calibration size, repetition, confidence level, and initial predictions, we perform the following conformalization procedures and assess their average population coverage on a separate test set of $10^6$ observations. 
The methods compared for conformalization are:
\begin{itemize}
    \item GPD profile: $\hat{C}^e(\bm{x})$ from \cref{e:extremePI}, using the GPD profile-likelihood CI for $\hat{q}^e_\alpha$.
    \item GPD bootstrap: $\hat{C}^e(\bm{x})$, using the GPD nonparametric bootstrap percentile CI for $\hat{q}^e_\alpha$.
    \item GPD delta: $\hat{C}^e(\bm{x})$, using the GPD delta-method CI for $\hat{q}^e_\alpha$.
    \item GPD simple: $\hat{C}^e(\bm{x})$, but using the simple GPD $(1-\alpha)$-quantile estimate of the calibration scores instead of the endpoint of a CI for $\hat{q}^e_\alpha$.%
    \item Classical: Single-sided version of the classical split conformalized quantile regression (still using the extreme quantile predictions $\hat{Q}^e_{1-\alpha}(\cdot)$, for a fair comparison). %
\end{itemize}
Each of the GPD-based procedures use the empirical $0.95$-quantile as the threshold $u$.

\subsection{Coverage results}\label{ss:simresults}

\begin{figure}[tb]
\centering
\includegraphics[width=\textwidth]{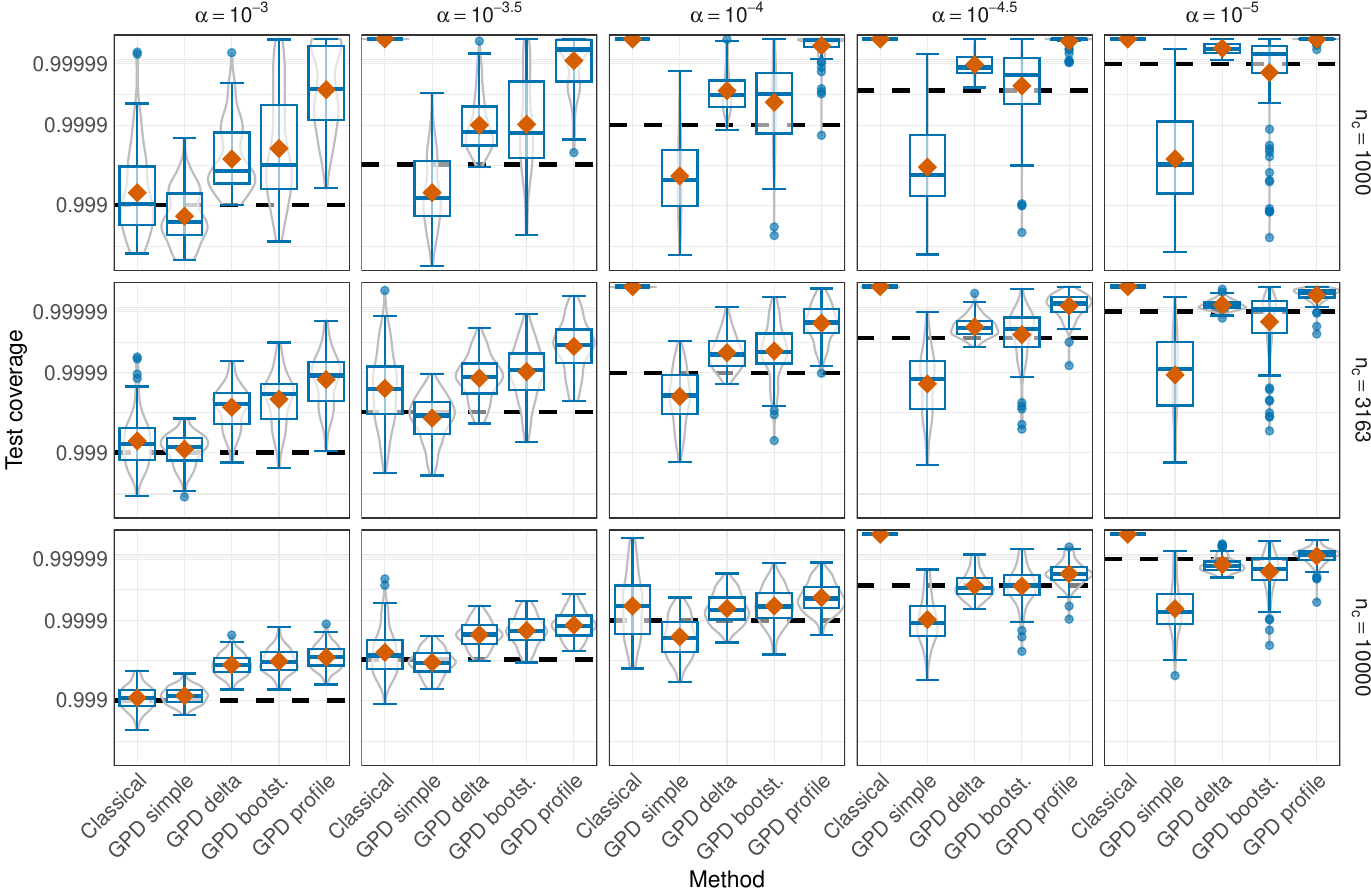}%
\caption{Boxplots of the test coverage probability of the quantile PIs (analytically computed for a grid of test observation and averaged over $\mathcal{X}$) for different conformalization methods, conformal confidence values $1-\alpha$ (columns, labelled with $\alpha$), and calibration sample sizes $n_c$ (rows), for the Student $t$ distributed noise and quantile ground-truth predictions.}%
\label{f:tsimtcal}
\end{figure}

\Cref{f:tsimtcal} shows the distribution of the computed test coverage for each considered conformalization method, confidence level, and calibration size, for the Student $t$ noise and ground truth original predictions. 
The chosen confidence levels are particularly large relative to the size of the calibration sets. 
Hence, for most scenarios, $\alpha<1/(n_c+1)$. 
In those cases, the classical conformalization method always yields trivial infinite intervals.
They have a coverage of $1$, but are uninformative and of no practical use. 
On the other hand, the other methods, relying on peaks over threshold extrapolation instead of empirical quantiles, are able to yield finite PIs, even when $\alpha\ll 1/(n_c+1)$. 

The simple GPD estimates of the calibration-score $(1-\alpha)$-quantile do not seem to provide sufficient coverage with small calibration samples and for the larger confidence levels, likely due to the GPD estimation error or approximation biases. 
The other three methods, relying on the confidence intervals for the score quantiles (and \cref{prop:CIcovr}), achieve much better coverage and, in most cases, satisfy \cref{e:mcov} as their coverage is larger than $1-\alpha$ on average.
However, those GPD CI-based methods seem consistently overconservative for lower confidence levels. 

In general, the profile likelihood method seems the most conservative, compared to the nonparametric bootstrap and delta method alternatives, as anticipated. It also satisfies the coverage guarantee in all scenarios. 
Its downside is the numerical difficulty, described in \cref{ss:extrconform}.
With the implementation at hand, this issue arose, in the worst case, in $85\%$ of the repetitions for $n_c=1000$ and the lowest $\alpha$ value, but quickly decreased to, at most, $2\%$ for $n_c= 3163$ and $0\%$ for $n_c= 10000$. 
This instability is understandable in the former truly extreme case, as the confidence level is more than two orders of magnitude larger than the level for which PIs are obtainable with the classical conformal method and as the likelihood only relies on $50$ observations to estimate a $(1-5\cdot 10^{-6})$-confidence interval for a $(1-5\cdot 10^{-6})$-quantile. 
However, this instability issue appears to often be resolved by slightly varying the choice of the GPD threshold, or of the~$\alpha_1$ and~$\alpha_2$ split.

The bootstrap and delta-method approaches seem less overconservative for the more moderate $\alpha$ values, but slightly undercover in the scenarios with the lowest $\alpha$ values. Nevertheless, they still significantly outperform the simple GPD approach and the infinite classical PIs. Contrary to the profile likelihood approach, the bootstrap method never fails to provide finite estimates. On the other hand, the delta-method approach also suffers from stability issues with small calibration sizes, regardless of the confidence level.

\Cref{f:tsimNcal} in \cref{apx:figs} shows the same coverage distribution when the data-generating process has light-tailed Gaussian noise, instead of heavy-tailed Student $t_4$ noise. 
In comparison, all methods tend to result in significantly more conservative intervals in terms of the coverage. 
In particular, all three GPD CI-based approaches always result in more coverage than necessary.

\begin{figure}[tb]
\centering
\includegraphics[width=\textwidth]{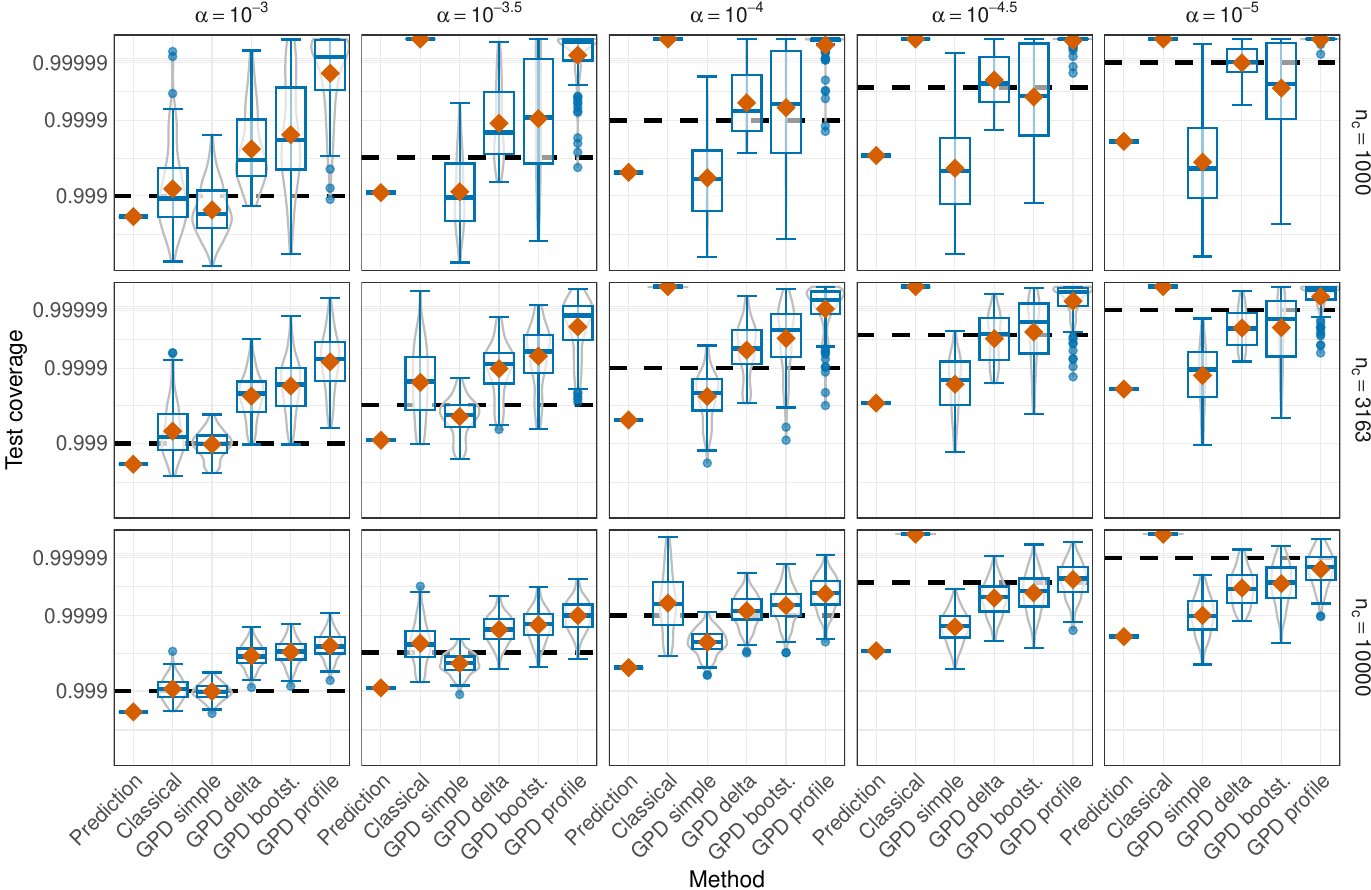}%
\caption{Boxplots of the test coverage probability of the quantile PIs (analytically computed for a grid of test observation and averaged over $\mathcal{X}$) for different conformalization methods, conformal confidence values $1-\alpha$ (columns, labelled with $\alpha$), and calibration sample sizes $n_c$ (rows), for the Student $t$ distributed noise and EQRN predictions.}%
\label{f:EQRNsimtcal}
\end{figure}

\Cref{f:EQRNsimtcal} shows the coverage distributions for the EQRN predictions and Student $t_4$ noise. 
We observe that, although being accurate predictions of the conditional quantile, in terms of integrated squared error \citep{Pasche2024}, the EQRN predictions, here, undercover when considered as a PI endpoint.
Note that even very accurate quantile predictions are still likely to lead to undercoverage, as a local quantile underestimation typically leads to a larger coverage loss than the coverage gain from a local overestimation of the same amplitude, due to the generally decreasing probability density in the tails.
The conformalization results closely match those for the ground-truth quantile predictions, although the coverage seems smaller for the largest confidence levels, for all methods. The scenario with $\alpha=10^{-5}$ and the largest sample size is the only one for which the GPD profile approach seems to slightly undercover. All the other finite alternatives also undercover for this largest confidence level. 
The CI-based extreme conformal prediction methods all outperform the original EQRN prediction in all scenarios.

\Cref{f:linsimtcal} in \cref{apx:figs} shows the same coverage distributions when using the poorly-fitting linear GPD quantile predictions, described in \cref{ss:simsetup}, instead of the EQRN model. 
The base predictions severely undercover, when used as PIs. 
We observe that, even with poor quantile modelling choices, in all considered scenarios, our extreme conformalization approach results in a mean test coverage always greater than the desired levels, in this case for all three GPD CI-based variants. 
Interestingly, the intervals seem overall more marginally conservative than with the EQRN base predictions. 
Thus, even with this choice of a poorly-fit base model, undoubtingly worsening local adaptivity and coverage, our method still results in nontrivial PIs consistently achieving sufficient marginal coverage.

As a takeaway, our practical recommendation for conservative and informative high-confidence PIs is to use the profile-likelihood version of our method.
In case it suffers from numerical issues and varying the GPD threshold or the $\alpha_1$ and $\alpha_2$ split slightly does not solve them (or if those are desired fixed), the bootstrap-based CI can be used instead.
This combination results in the profile-likelihood conservativeness, in most scenarios, and avoids the cases of potential numerical difficulties by using the bootstrap estimate, which is still conservative enough in the majority of scenarios. 
We call this method the ``GPD safeprofile'' PI. 
Alternatively, considering the maximum of the bootstrap and delta-method PI endpoints as a replacement in unstable profile likelihood situations could be more conservative but might be less stable.

\section{Application to flood risk forecasting}\label{s:appl}

\subsection{Description and aim}

Flooding is one of the most impactful natural hazards in terms of infrastructure and economic damage, and of the endangerment of human lives. 
Methods from extreme value theory have proven successful for assessing flood risk and providing reliable worst-case scenarios
\citep[e.g.,][]{kat2002, kee2009, ASADI2018182, eng2018, Engelke2020}.

In Switzerland, the Federal Office for the Environment (FOEN) monitors the river flow with numerous gauging stations throughout the river network. In its capital city, Bern, extreme water flow events of the Aare river led to several major floods, causing some of the most severe infrastructural and economic flooding damages recorded in the country. 
The main driver of strong water-flow events is the cumulative amount of upstream precipitation. 
In this study, we rely on the average daily discharge measures of the Aare river (in $\text{m}^3\text{s}^{-1}$) provided by the FOEN\footnote{\url{https://www.hydrodaten.admin.ch/}.}, and on recordings of daily precipitation (in $\text{mm}$) at various meteorological stations, obtained from MeteoSwiss\footnote{\url{https://opendatadocs.meteoswiss.ch/}.}. 
This version of the dataset was preprocessed and analysed in previous studies~\citep{Pasche2021,Pasche2024}.

With our proposed extreme conformal approach, we aim to provide high-confidence one-day-ahead interval forecasts of the conditional range for water flow. 
We rely on several extreme quantile regression models pretrained to forecast the one-day-ahead extreme quantiles of the Aare water flow in Bern, given observations of upstream precipitation at six locations in Bern's water catchment and of the average daily water flow at an upstream gauging station, during the previous 10 days.
\Cref{f:chstations} in \cref{apx:figs} shows the location of those meteorological and gauging stations. 

\subsection{Methodology}

The river data exhibits a strong seasonal pattern with more water flow, and with more frequent and intense extreme events, during the late spring and summer months, due to snowmelt and heavy precipitation. 
This seasonality is likely to propagate to the residual nonconformity scores, violating the exchangeability assumption underlying classical conformal prediction.
To account for it, we use the weighted variation of our extreme conformalization approach, discussed in \cref{ss:extensions}.
As a general procedure to account for distribution drift, \citet{Barber2023nonstatConform} suggest choosing large weights for recent calibration observations, close in time to the test period, and having weights decay for earlier observations, either exponentially or as a cutoff.
Here, we make use of the inherent periodicity of the seasonal river discharge behaviour, by matching it with sinusoidal weights. 
Each year is partitioned into 24 roughly equal seasonal blocks of 15 to 16 days each. 
Let $B(i)\in\{1,\ldots,24\}$ denote the block in which an observation indexed by $i$ falls into. 
For each block $b=1,\ldots,24$, $\hat{q}^e_\alpha(b)$~is estimated as the upper-endpoint of the appropriate extreme score quantile CI, as described in \cref{ss:extrconform}, but using the weighted GPD likelihood in \cref{e:wgpdll}, with calibration sample weights 
\begin{equation}\label{e:applweights}
    w_i = \cos\left[\frac{2\pi}{24} \{B(i) - b\}\right] + 1, \quad i=1,\ldots,n_c.
\end{equation}
This choice of weights gives the highest importance to calibration observations in the same seasonal block $b$ as the test point, and decreases the importance of observations in blocks further away in the year, with a yearly periodicity.
For the estimation of $\hat{q}^e_\alpha(b)$, we use the GPD safeprofile CI procedure recommended in \cref{s:sim}, although almost no numerical issues were encountered with the profile likelihood.

Our final one-day-ahead extreme PI for the average daily discharge, during seasonal block $b=1,\ldots,24$ of the year, given past observations of upstream precipitation and water flow $\Xtest=\bm{x}$, is then
$\hat{C}^e(\bm{x}) = \left(0,\; \hat{Q}^e_{1-\alpha}(\bm{x}) + \hat{q}^e_\alpha(b)\right]$.
We compare this PI to the unconformalized quantile predictions $\hat{Q}^e_{1-\alpha}(\bm{x})$, and to a single-sided (see \cref{ss:singlesidedPI}) and weighted \citep{Barber2023nonstatConform} variation of the classical split conformalized quantile regression PI $\hat{C}(\bm{x})$ \citep{ConformalizedQR}, using the same sinusoidal weights from \cref{e:applweights}.

Apart from its seasonality, the data is also sequentially dependent. However, the residual dependence between the scores $S_i$ is likely weak and short-term, which should not significantly affect the marginal coverage guarantee of the conformal PIs~\citep{OliveiraConformNonexchError}.

We consider several choices of extreme quantile regression models for $\hat{Q}^e_{1-\alpha}(\cdot)$: EQRN, GBEX~\citep{gbex}, EGAM~\citep{ExGAM2}, and EXQAR~\citep{EXQAR}. We also consider the constant unconditional GPD quantile estimates as a comparative baseline. 
We emphasise results using the recurrent version of EQRN, which is specifically designed for sequential dependence, and seems to fit the data best~\citep{Pasche2024}.

The quantile models were pretrained on data from 1939 to 1951. 
They were all fine-tuned with a grid search for hyperparameter selection. 
We use data from 1951 to 1999 for calibration and testing, i.e.\ 48 years of daily data. 
The observations after 1999 are not considered, due to a major distribution shift\footnote{See the flood report of the FOEN at \url{https://www.hydrodaten.admin.ch/en/2135.html}.}.
We choose the first 10 years as the default calibration set in the first part of the analysis, but vary its size from 3 to 15 years in the second part, and use the rest for estimating PI coverage empirically. 
We use multiples of complete years to keep a seasonal balance in the calibration and test sets.

\subsection{Results}

\begin{figure}[tb]
\centering
\includegraphics[width=\textwidth]{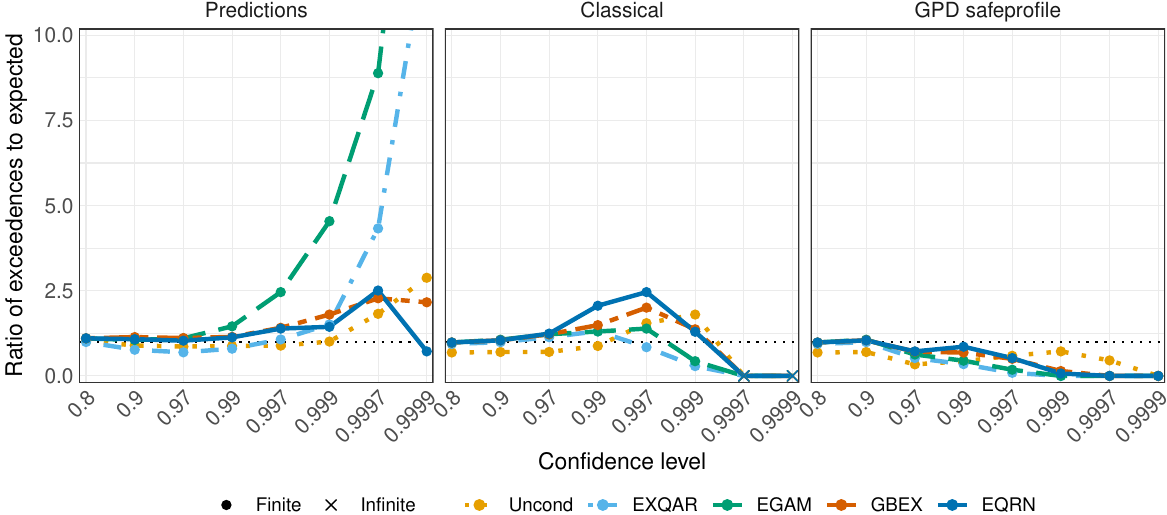}%
\caption{Number of observations exceeding the PIs during the test period as a ratio to the expected number of exceedances for different confidence levels and each pretrained prediction model, for the original predictions (left), the classical conformalization (center), and the GPD safeprofile method (right).}%
\label{f:CHcal10allratio}
\end{figure}

\Cref{f:CHcal10allratio} compares the number of observations exceeding the PIs from each method during the test period, using predictions of the different pretrained models, for a range of moderate to extreme confidence levels. 
The number of observations expected to exceed the PIs during the 38-year test period varies from \numprint{2776}, for $1-\alpha=0.8$, to only $1.4$, for the largest level $1-\alpha=0.9999$. 
Using the original model predictions as PIs leads, in most cases, to undercoverage. Although the best-performing predictions seem to vary around the target coverage, they fail to provide satisfactory coverage consistently.
The classical conformalization seems effective for the lowest two confidence levels but worsens the coverage for the following moderately extreme levels, compared to the initial predictions. 
At the two largest levels, the classical method yields uninformative infinite PIs. 
The extreme conformalization method yields finite PIs with significantly better coverage for confidence levels above $0.95$, for which the approaches differ. 
The PI coverage consistently strictly satisfies the target confidence levels for each initial prediction model. 
The profile likelihood procedure was stable for almost all models, confidence levels and seasonal blocks.
A numerical instability only occurred once, for the unconditional predictions at the largest confidence level, during a single one of the seasonal blocks.

\begin{figure}[tbhp]
\centering
\includegraphics[width=0.9\textwidth]{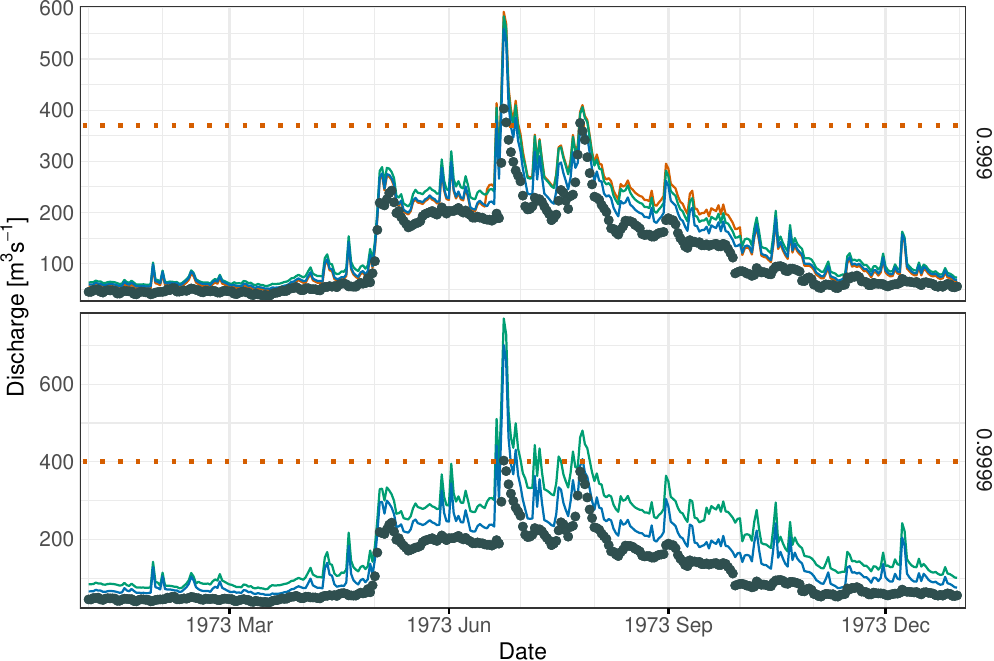}%
\caption{Original EQRN prediction (blue), classical conformal PI (red), and extreme conformal PI (green), at confidence levels $0.999$ (top panel) and $0.9999$ (bottom panel), during one of the test years. The classical conformal PI is infinite at level $0.9999$. The observations (points) and the unconditional GPD $(1-\alpha)$-quantile estimates (dotted lines) are also shown.}%
\label{f:CHcal10preddQp9999}
\end{figure}

\Cref{f:CHcal10preddQp9999} shows the initial EQRN predictions, which seem to fit the data variation best, and their conformalized PI endpoints for two of the considered confidence levels, including the largest, during a test year. 
At $1-\alpha=0.999$, even though it has more marginal coverage, the extreme PIs are not excessively larger than the classical PIs. 
The extreme intervals are even tighter during some of the seasonal blocks.
At the largest level, the infinite classical-method PIs are uninformative. 
On the other hand, the extreme PIs, satisfying the desired test coverage, are again not overly large compared to the data variation, the original predictions, and the unconditional quantile estimates. 
Extreme conformal corrections for the EQRN predictions at the other unshown levels are smaller in magnitude. 
During the considered year, the original predictions at the largest confidence level are exceeded once, on 25th July 1973. This exceedance is covered by the conformalized extreme PI.
Finally, we observed the localised adaptivity of weighted conformalization, as, for example, the conformal correction is larger during late summer than during winter for both methods and confidence levels.

\begin{figure}[tbhp]
\centering
\includegraphics[width=0.95\textwidth]{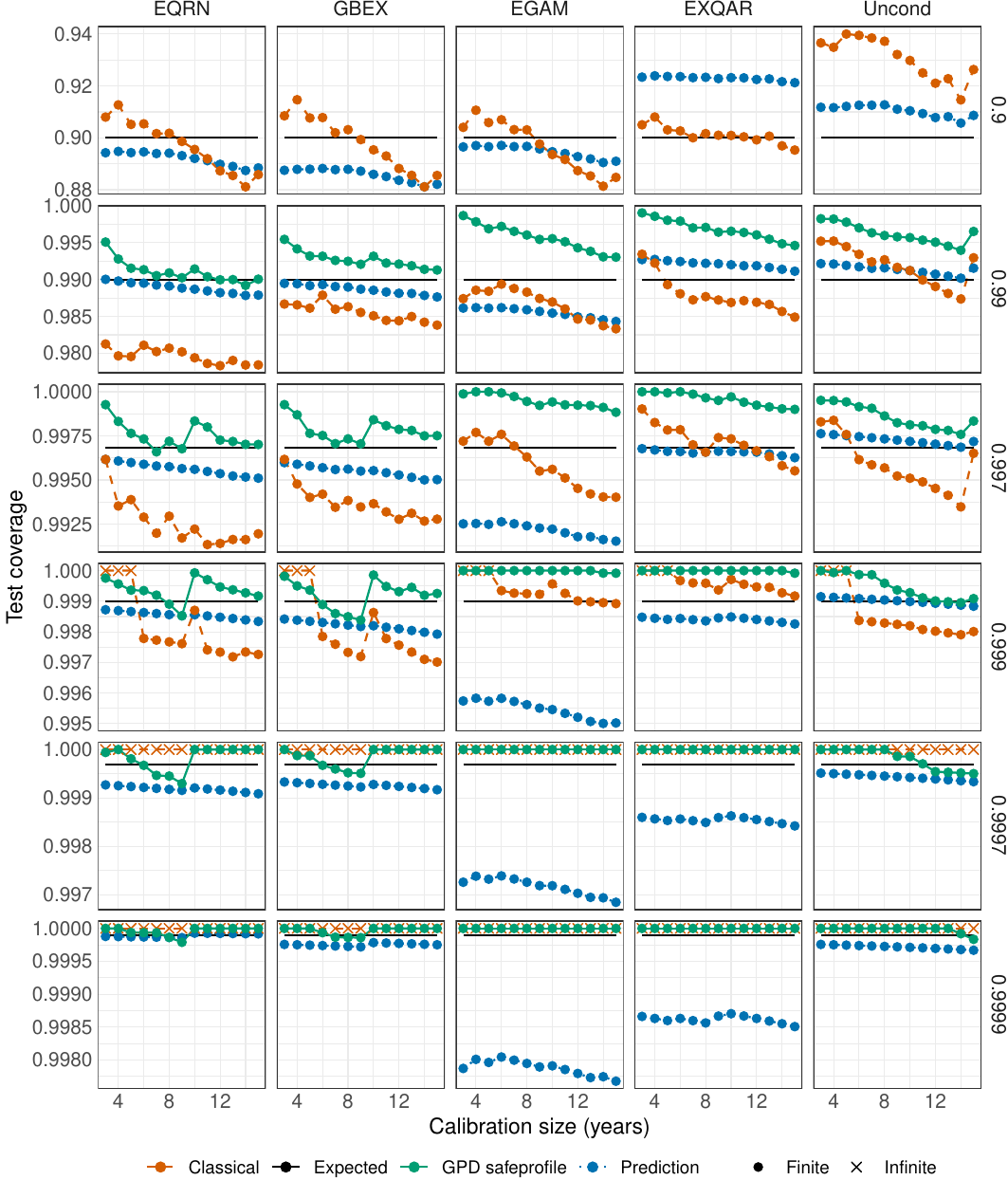}%
\caption{Empirical test coverage of the prediction intervals for a range of calibration-set sizes, for each conformalization method, pretrained model (columns), and confidence level $1-\alpha$ (rows). For $1-\alpha<0.95$, corresponding to the quantile level of the GPD threshold $u$, the GPD approach coincides with the classical.}%
\label{f:CHncCovr}
\end{figure}

\Cref{f:CHncCovr} shows the evolution of the test coverage with the calibration size, for all predictions, methods, and different confidence levels. 
We observe that the extreme PIs significantly outperform the classical conformalization in terms of empirical test coverage, for all relevant levels.
It always provides informative finite PIs, contrary to the classical approach that yields infinite PIs for calibration sizes up to 5 years with $1-\alpha=0.999$, and for all sizes at the two largest levels. 
The extreme PI has valid coverage in almost all combinations, including when the classical approach and/or original predictions significantly undercover. 
In the few undercovered situations, its coverage is closer to the target $1-\alpha$ than both the original predictions and the classical approach.
For some of the lower levels, there seems to be a pattern of decreasing coverage with increasing calibration size for the conformalized PIs. 
The largest discharge events happening in years one, three and 15 of the chosen calibration period is a likely explanation for this decrease in between, although there might also be a small effect from the decreasing test size, as observed with the coverage of the unconformalized predictions.

\clearpage

\section{Conclusion}\label{s:concl}

We propose a conformalization method which relies on extreme value statistics to provide conservative and nondegenerate prediction intervals for reliable risk assessment of high-impact events under extreme-confidence requirements. 
The novel method uses the well-studied peaks-over-threshold approach, which leverages the generalized Pareto distribution to extrapolate the necessary conformal correction from the calibration data to the required extreme confidence levels.
It uses a conservative confidence interval solution for robustness against estimation and approximation biases.

In the simulation study and the application to forecasting river flow, at large confidence levels, our extreme conformal prediction method consistently provides PIs with significantly better coverage properties than both the original predictions and classical conformalization.
For the largest levels, classical approaches result in infinite (or undefined) PIs, which are of no practical use. 
On the other hand, our recommended method yields informative finite intervals that consistently achieve the desired coverage for confidence levels up to several orders of magnitude larger than what is feasible with classical methods. %
The weighted version of our approach can account for nonstationary data drifts, such as the seasonal behaviour of the river flow.
Importantly, our extreme conformal prediction method can be used in combination with any extreme quantile regression model, including black-box machine learning methods without known asymptotic guarantees.

One downside of our approach is its potential overconservativeness in certain scenarios, e.g., for moderately extreme confidence levels and some lighter-tailed data distributions.
This is, at least in part, a consequence of our conservative CI-based solution, used to circumvent possible estimation and approximation biases in the GPD estimation.
Moreover, peaks-over-threshold approaches rely on asymptotic properties. 
Thus, our approach lacks the finite-sample guarantees of classical conformal prediction.
However, this is an unavoidable tradeoff for the ability to extrapolate beyond the moderate levels for which the classical empirical-quantile-based methods are feasible. 
In fact, obtaining finite-sample bounds for quantiles extrapolated beyond the data range is not theoretically possible without additional assumptions on the data distribution~\citep{BoucheronThomasConcentration,ThomasThesisEVTConcentr,LhautEVTConcentr}. 
A similar tradeoff exists in conformal prediction for conditional coverage, which is impossible to ensure on finite samples, resulting in the formulation of asymptotic conditional coverage guarantees as an alternative \citep{LeiWassermanConformalLocalAsymptotic}.

This work establishes a first method for extreme-confidence PIs, as an extension of the well-established conformal regression framework. 
As our experiments mainly focused on the split-conformal approach, due to its computational efficiency and practical popularity, further work is possible with alternative or more specialised conformal approaches.
One direction is to enhance the efficiency of data usage, using, for example, the full conformal version of our approach, described in \cref{apx:extensionsdetails}, or potential variants of jackknife \citep{barber2021predictive,steinberger2016leave}, infinitesimal jackknife \citep{alaa2020discriminative}, or cross-validation \citep{vovk2015cross}. 
However, such approaches typically have a much greater computational cost, as they require refitting the predictive model many times. 
Whether the gain in statistical efficiency for our extreme conformal approach from using these alternative conformalization procedures is worth the extra computational cost remains an open question.
Other directions include further extensions to nonexchangeable data.
The weighted version of our extreme conformal approach seems suitable for nonstationary covariate drifts and shifts \citep[analogously to][]{Barber2023nonstatConform,TibshiraniCoformShift}.
However, alternative solutions might be necessary to account for other problematic scenarios, such as long-term dependence or drifts in the conditional distribution of the response given the covariates. 
Finally, approaches other than conformalization could also warrant investigations, such as building PIs based on the so-called high-quality criterion and using deep learning \citep{DLpredintervals,khosravi2011comprehensive,ChenLam21PI}. 
How to incorporate extreme value statistics to extrapolate prediction intervals to cover high-impact events in these methodologies appears to be largely open.

\newpage

\section*{Supplementary material}

\subsection*{Implementation and reproducibility}
To facilitate its practical use, the proposed extreme conformal procedure, including its weighted variant for nonstationary data, is implemented as an open-source R package, available at \url{https://github.com/opasche/ExtremeConformal}. 
Furthermore, a new versatile and efficient algorithm for estimating profile-likelihood confidence intervals for extreme quantiles (and return levels) is used as a dependency and is available as a separate R package at \url{https://github.com/opasche/ExtremeCI}.
The code and data, with detailed instructions for reproducing the results presented in this paper, are available at \url{https://github.com/opasche/Reprod_ExtremeConformalPred}.

\section*{Declarations}

\subsection*{Acknowledgements}
This research project was conducted while the first author, O.~C.~Pasche, was a visiting scholar at the Department of Industrial Engineering and Operations Research, at Columbia University. He thanks the department and the university for their hospitality during this period. 
We also thank the reviewers and the associate editor for their valuable comments. 

\subsection*{Funding}
O.~C.~Pasche and S.~Engelke were supported by the Swiss National Science Foundation Eccellenza Grant 186858. 
H.~Lam was supported by the InnoHK initiative of the Innovation and Technology Commission of the Hong Kong Special Administrative Region Government, Laboratory for AI-Powered Financial Technologies, and the Columbia Innovation Hub Award.

\subsection*{Availability of supporting data}
In the Application to flood risk forecasting, we use river discharge and precipitation data recorded in Switzerland between 1930 and 1999, in the Rhine and Aare basins. The precipitation records can be freely obtained from MeteoSwiss, on \url{https://opendatadocs.meteoswiss.ch/}, and the discharge records from the Swiss Federal Office for the Environment (FOEN), on \url{https://www.bafu.admin.ch/bafu/en/home/topics/water/data-and-maps/water-monitoring-data/hydrological-data-service-for-watercourses-and-lakes.html}. 

\subsection*{Published article}
This document is the preprint of an article in press for publication in the special issue `Bridging Heavy Tails and Artificial Intelligence' of \emph{Extremes}~\citep{Pasche26ExtrConform}, with \doi{10.1007/s10687-026-00536-9}.
When citing this work, please refer to the published version.

\clearpage

\bibliographystyle{apalike_OCP}%
\renewcommand{\bibfont}{\footnotesize}
\setlength{\bibsep}{0.45\bibsep}

\bibliography{OCP_bib}

\newpage

\appendix

\begin{center}
{\LARGE\bf 
APPENDIX
}
\end{center}

\vspace{10mm}

\setcounter{section}{0}
\setcounter{subsection}{0}
\setcounter{equation}{0}
\setcounter{figure}{0}
\setcounter{table}{0}

\renewcommand{\thesection}{A.\arabic{section}}
\renewcommand{\thesubsection}{A.\arabic{section}.\arabic{subsection}}
\renewcommand{\theequation}{A.\arabic{equation}}
\renewcommand{\thefigure}{A.\arabic{figure}}
\renewcommand{\thetable}{A.\arabic{table}}

\renewcommand{\theHsection}{A.\thesection}
\renewcommand{\theHsubsection}{A.\thesubsection}
\renewcommand{\theHequation}{A.\theequation}
\renewcommand{\theHfigure}{A.\thefigure}
\renewcommand{\theHtable}{A.\thetable}

\section{Extensions to other conformal approaches}\label{apx:extensionsdetails}

This section describes in more detail the extensions of our proposed extreme conformal method to alternative conformal procedures discussed in \cref{ss:extensions}.

As explained in the main text, the extension to different base predictive models and scores is the most straightforward, as our extreme conformalization described in \cref{ss:extrconform} is agnostic to their definition.
This includes the classical split-conformal approach that, instead of the quantile predictions $\hat{Q}_{\alpha/2}(\bm{x}), \hat{Q}_{1-\alpha/2}(\bm{x})$, relies on a conditional-mean base regression model $\hat{\mu}(\bm{x})$, and the residuals $s(\bm{x},y):= \abs{y - \hat{\mu}(\bm{x})}$ as nonconformity score \citep{SplitConformalMargProof,papadopoulos2008inductive,Lei18ConformalPred}. 
The procedure from \cref{ss:extrconform} would then be performed with these alternative scores, resulting in the fixed-length extreme conformal PIs $\hat{C}^e(\bm{x}) = \left[\hat{\mu}(\bm{x}) - \hat{q}^e_\alpha, \hat{\mu}(\bm{x}) + \hat{q}^e_\alpha\right]$. 
If a residual-dispersion estimate $\hat{\sigma}(\bm{x})$ is also available, for instance, from a heteroscedastic regression model or a Bayesian approach, using the scaled residuals $s(\bm{x},y):= \abs{y - \hat{\mu}(\bm{x})}/\hat{\sigma}(\bm{x})$ as nonconformity scores \citep{papadopoulos2008normalizedCoform,papadopoulos2011regressionknn,Lei18ConformalPred} would result in the varying-length extreme PIs $\hat{C}^e(\bm{x}) = \left[\hat{\mu}(\bm{x}) - \hat{q}^e_\alpha \hat{\sigma}(\bm{x}), \hat{\mu}(\bm{x}) + \hat{q}^e_\alpha \hat{\sigma}(\bm{x})\right]$. 
Although the latter doesn't require extreme quantile regression to yield varying-length PIs, it tends to underestimate conditional variability and yield less adaptive predictions~\citep{ConformalizedQR}.

In full-conformal procedures, the data is not split into training and calibration sets. 
Instead, given a training set $\{(\bm{X}_{i}, Y_{i})\}_{i=1}^n$ and the test covariates $\Xtest$, the PI $\hat{C}(\Xtest)$ for $\Ytest$ is constructed by refitting the base model $\hat{f}$ (e.g., $\hat{Q}_{1-\alpha}$ or $\hat{\mu}$) on $\{(\bm{X}_{i}, Y_{i})\}_{i=1}^n \cup \{(\Xtest, y)\}$, for a dense grid of values $y$ in the value space of $Y$. Each fit is denoted by $\hat{f}^{y}$. 
The desired nonconformity scores are then defined, for each $y$, as $S_i(y):=s(\bm{X}_i, Y_i; \hat{f}^{y})$, $i=1,\ldots,n$ \citep{vovk1999ml,vovk2005algorithmic,shafervovk2008tutorial}.
Consequently, for a full-conformal variant, our proposed procedure from \cref{ss:extrconform} should be performed, for each $y$, on the scores $\{S_i(y)\}_{i=1}^n$, to obtain a $\hat{q}^e_\alpha(y)$. 
The resulting PI is then 
$\hat{C}^e(\Xtest) = \{y : s(\Xtest, y; \hat{f}^{y}) \leq \hat{q}^e_\alpha(y)\}$.
Although the latter makes a more efficient use of the data than the split variant, it is extremely computationally costly.
The analogous extension to the middle-ground $k$-fold approaches, such as Jackknife+/CV+ \citep{barber2021predictive} and cross-conformal prediction \citep{vovk2015cross}, would also suffer, less extremely, from similar refitting expensiveness.

\section{Proof of Proposition~\ref{prop:CIcovr}}\label{apx:proof_covr}

\begin{proof}
Let $[L_Q,U_Q]$ be the $(1-\alpha_2)$-confidence interval for $q:={F}^{-1}_{S}(1-\alpha_1)$. 
Then, by assumption, 
$$\Prob(q\leq U_Q)\geq \Prob(L_Q\leq q\leq U_Q)\geq 1-\alpha_2, \quad\text{ and }\quad \Prob\left(\Stest \leq q\right)\geq 1-\alpha_1.$$
As $\Stest$ and $U_Q$ are independent, 
\begin{equation*}
\begin{split}
\Prob\left(\Stest \leq U_Q\right) &= \Prob\left(\Stest-q \leq U_Q-q\right)\geq\Prob\left(\{\Stest\leq q\} \cap \{U_Q\geq q\}\right) \\ %
&=\Prob\left(\Stest\leq q\right) \Prob\left(U_Q\geq q\right) \geq (1-\alpha_1)(1-\alpha_2)\geq1-\alpha.
\end{split}
\end{equation*}
By the definitions of $\hat{C}^e(\bm{x}) = \{y : s(\bm{x}, y) \leq U_Q\}$ and of $\Stest=s(\Xtest,\Ytest)$, the events\break{} \mbox{$\{\Ytest \in \hat{C}^e(\Xtest)\}$} and \mbox{$\{\Stest \leq U_Q\}$} are equivalent.
Therefore, 
$$\Prob\{\Ytest \in \hat{C}^e(\Xtest)\} = \Prob\left(\Stest \leq U_Q\right) \geq 1-\alpha.$$
\end{proof}

\section{Additional figures}\label{apx:figs}

\subsection{Simulation study}
\begin{figure}[bthp]
\centering
\includegraphics[width=\textwidth]{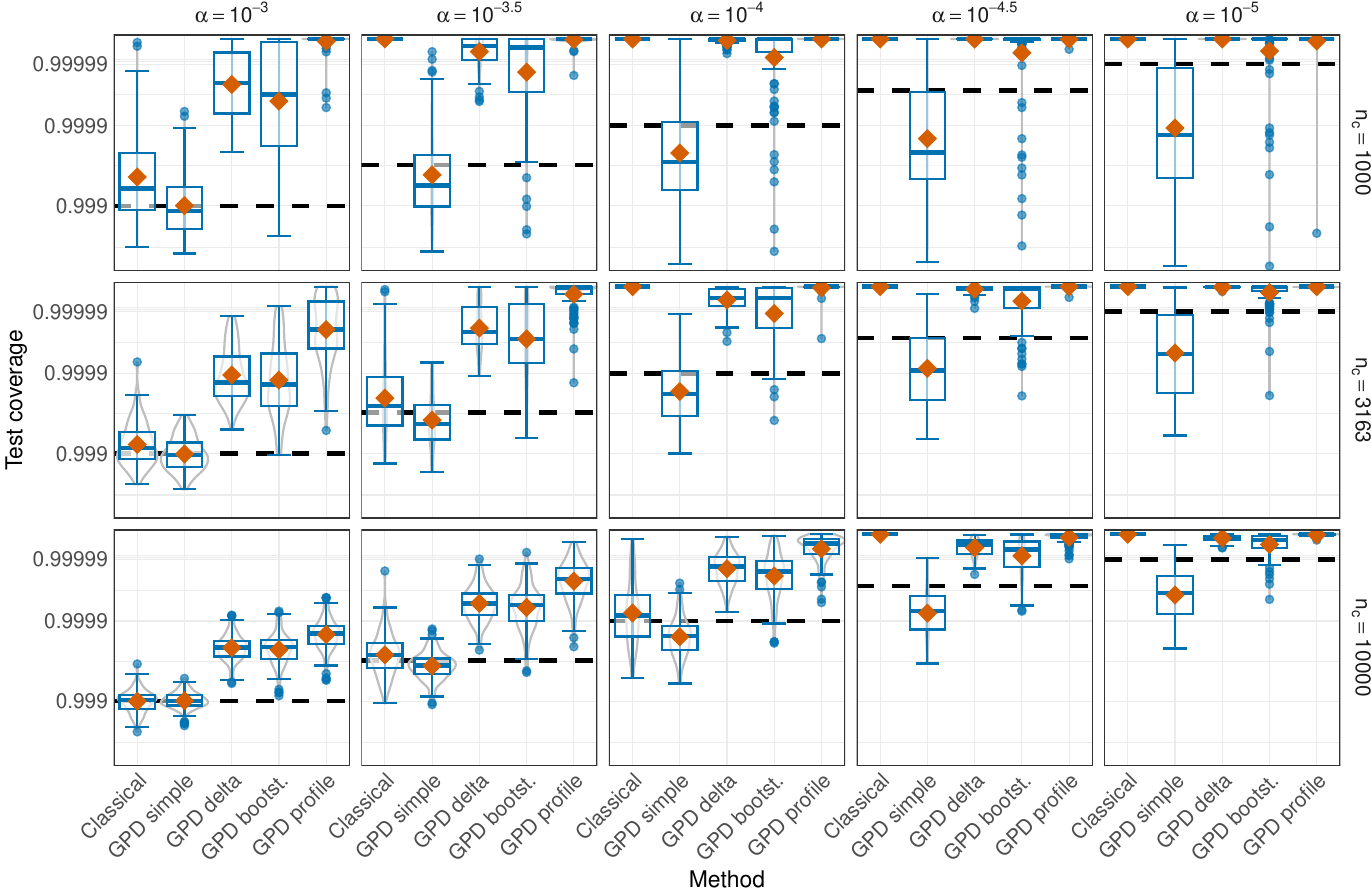}%
\caption{Boxplots of the test coverage probability of the quantile PIs (analytically computed for a grid of test observation and averaged over $\mathcal{X}$) for different conformalization methods, conformal confidence values $1-\alpha$ (columns, labelled with $\alpha$), and calibration sample sizes $n_c$ (rows), for the Gaussian distributed noise and quantile ground-truth predictions.}%
\label{f:tsimNcal}
\end{figure}

\begin{figure}[bthp]
\centering
\includegraphics[width=\textwidth]{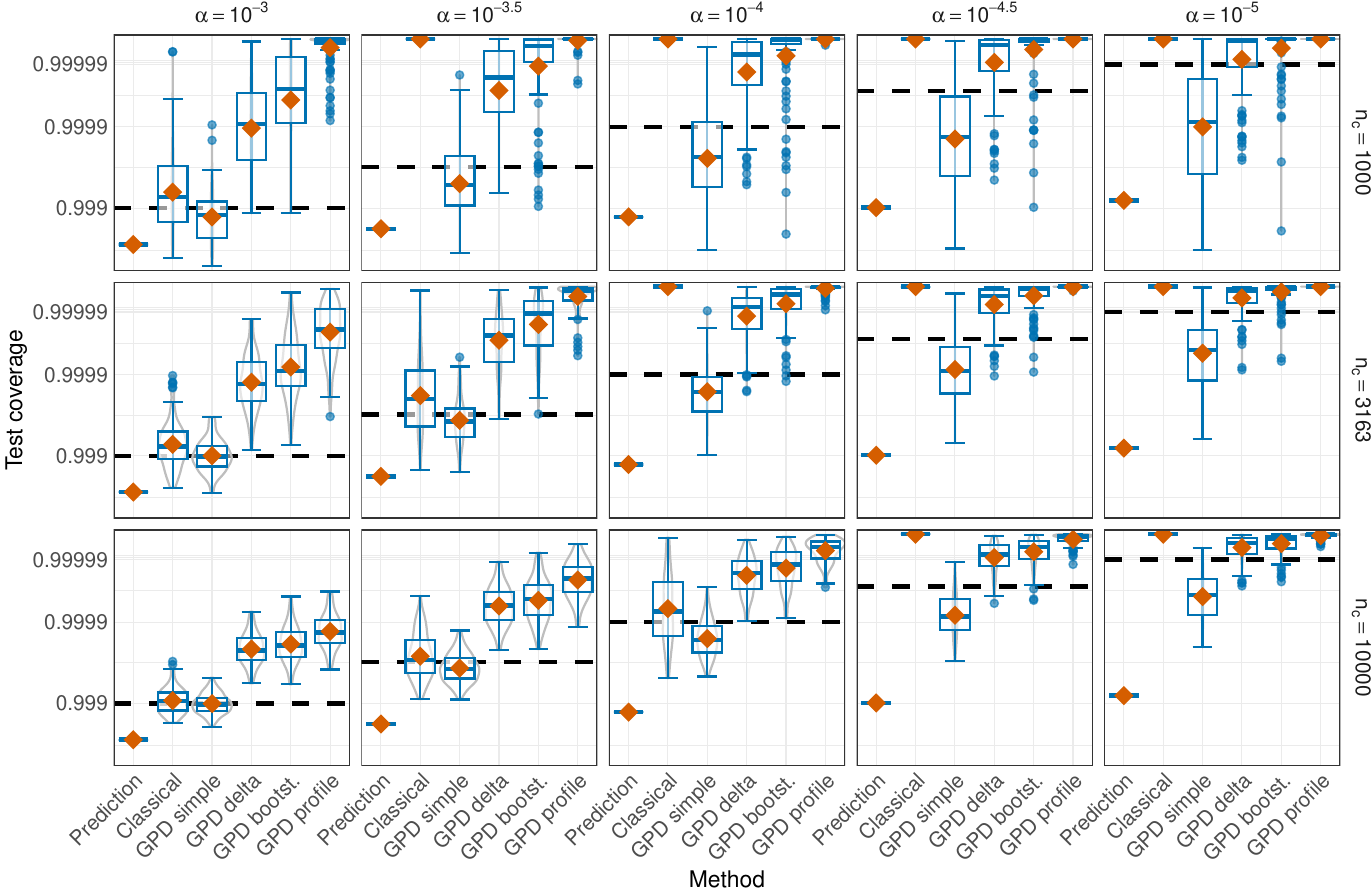}%
\caption{Boxplots of the test coverage probability of the quantile PIs (analytically computed for a grid of test observation and averaged over $\mathcal{X}$) for different conformalization methods, conformal confidence values $1-\alpha$ (columns, labelled with $\alpha$), and calibration sample sizes $n_c$ (rows), for the Student $t$ distributed noise and linear GPD quantile predictions.}%
\label{f:linsimtcal}
\end{figure}

\Cref{f:tsimNcal} shows the distribution of the computed test coverage for each considered conformalization method, confidence level, and calibration size, for the light-tailed Gaussian-noise data and the ground truth as base predictions. 
\Cref{f:linsimtcal} shows the same test-coverage distribution, for each considered conformalization method, confidence level, and calibration size, for the Student $t$ distributed noise and the linear GPD quantiles as the base predictions.

\clearpage

\subsection{Application to river-flow forecasts}
\begin{figure}[tbh]
\centering
\includegraphics[width=0.85\textwidth]{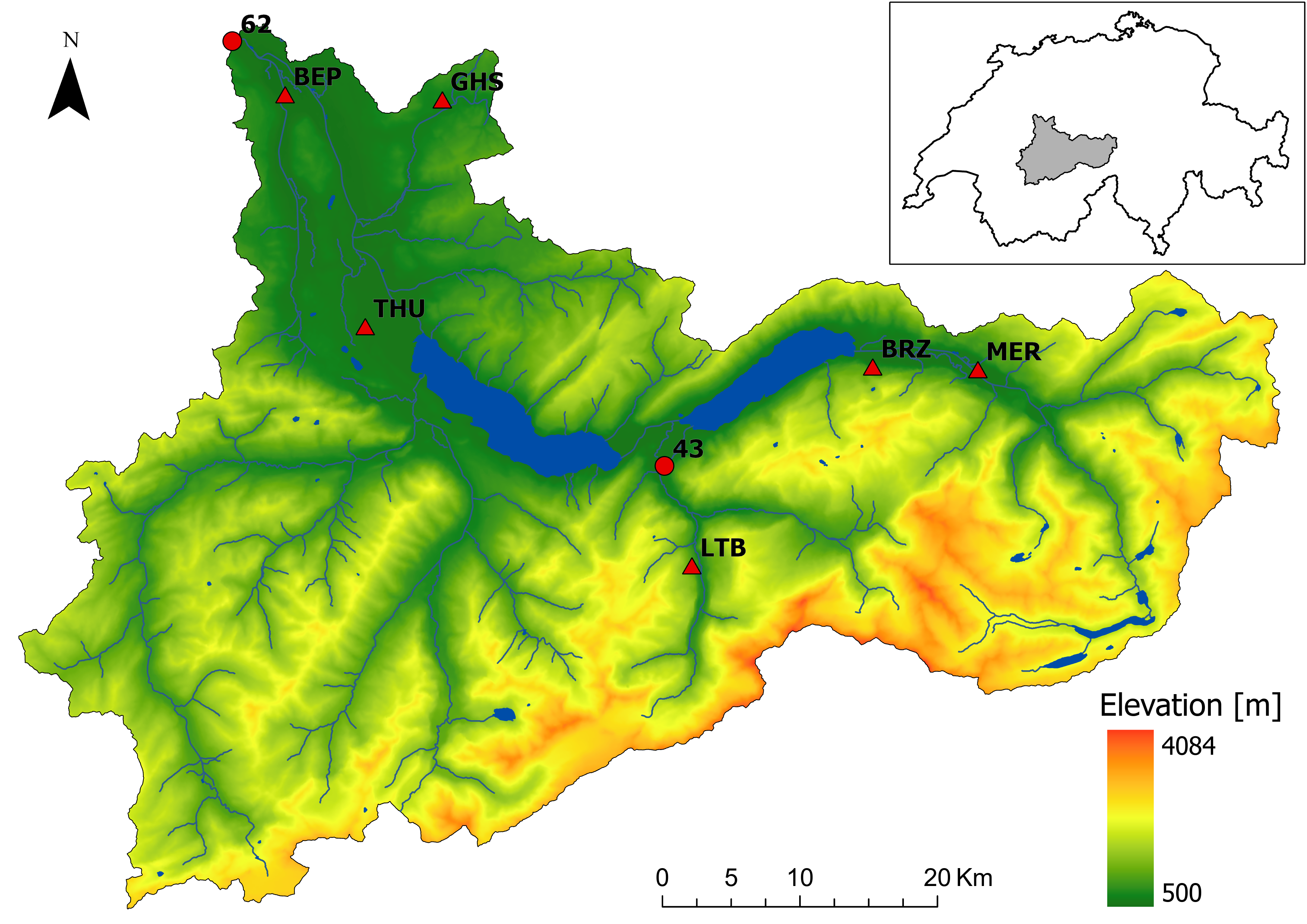}%
\caption{Topographic map of water catchment of the gauging station in Bern--Sch\"onau (62) on the Aare in Switzerland. Another upstream gauging station in Gsteig (43), on the L\"utschine river, and six meteorological stations with precipitation measurements (triangles) are also shown \citep[source:][]{Pasche2024}.}%
\label{f:chstations}
\end{figure}

\Cref{f:chstations} shows the locations of the meteorological and gauging stations corresponding to the variables used in the model forecasts. 

\end{document}